\documentclass[manuscript,screen,nonacm]{acmart}
\acmConference[]{} % need to include this to suppress addresses footer

\newcommand{\workshopname}{GenAICHI: CHI 2024 Workshop on Generative AI and HCI}

\newcommand{\licensedetails}{Licensed under a Creative Commons Attribution 4.0 International License (CC BY 4.0). Copyright remains with the author(s).}

\newcommand\extrafootertext[1]{% this command adds a non-numbered footnote
    \bgroup
    \renewcommand\thefootnote{\fnsymbol{footnote}}%
    \renewcommand\thempfootnote{\fnsymbol{mpfootnote}}%
    \footnotetext[0]{#1}%
    \egroup
}

\AtBeginDocument{ % setup headers and footers
    \fancypagestyle{firstpagestyle}{
        \fancyhf{}
        \fancyfoot[L]{\sffamily\footnotesize \workshopname}%
        \fancyfoot[C]{\sffamily\footnotesize \thepage}
    }
    \fancyhf{}
    \fancyhead[L]{\sffamily\footnotesize\shorttitle}
    \fancyhead[R]{\sffamily\footnotesize\shortauthors}
    \fancyfoot[L]{\sffamily\footnotesize\workshopname}%
    \fancyfoot[C]{\sffamily\footnotesize\thepage}
    \extrafootertext{\licensedetails}
}
\setcopyright{acmlicensed}
\usepackage{comment}
\usepackage{hyperref}

\usepackage[table,xcdraw]{}
\usepackage{float}
\usepackage{graphicx} 
\usepackage{multirow}
\usepackage{pifont}% http://ctan.org/pkg/pifont
\newcommand{\cmark}{\ding{51}}%
\newcommand{\xmark}{\ding{55}}%

\begin{document}
		
	\title{Seizing the Means of Production: Exploring the Landscape of Crafting, Adapting and Navigating Generative AI Models in the Visual Arts}
	% Make it Your Own: Towards Personal Generative AI through Model Crafting
	%Towards Personal Gen AI: Surveying the Landscape of Personalization for Generative AI
	
	\author{Ahmed M. Abuzuraiq}
	\email{aabuzura@sfu.ca}
	\orcid{0000-0002-3604-7623}
	\affiliation{
		\institution{Simon Fraser University}
		\country{Canada}
	}
	
	\author{Philippe Pasquier}
	\email{pasquier@sfu.ca}
	\orcid{0000-0001-8675-3561}
	\affiliation{
		\institution{Simon Fraser University}
		\country{Canada}
	}

	\renewcommand{\shortauthors}{Abuzuraiq and Pasquier}
	
	\begin{abstract}
		In this paper, we map out the landscape of options available to visual artists for creating personal artworks, including crafting, adapting and navigating deep generative models. Following that, we argue for revisiting model crafting, defined as the design and manipulation of generative models for creative goals, and motivate studying and designing for model crafting as a creative activity in its own right.
	\end{abstract}



	%\received{26 February 2024}
	%TODO \received[revised]{12 March 2009}
	%TODO \received[accepted]{5 June 2009}
	
	\maketitle
	
	\section{Introduction}
	Large-scale Text-to-image Generation Models (LTGMs), including Dalle-E~\cite{openai_2024_DALLE} and Stable Diffusion~\cite{podell_2023_SDXLImprovingLatent}, are trained on large datasets with many high-end GPUs. Users of these models can produce diverse and high-quality visuals through meticulously written text prompts. These models mark a significant shift from the era when artists used personally-trainable generative models like Generative Adversarial Networks (GANs) and Variational Autoencoders (VAEs). LTGMs have made the generation of visuals accessible to everyone, but this shift in attention has overshadowed the practice of "model crafting", whereas artists personalize their work by experimenting with training sets, model architectures, and hyperparameters in addition to combining, adapting and manipulating pre-trained models~\cite{broad_2021_Activedivergencegenerative}. Model crafting offered artists a sense of craftsmanship and ownership over the creative process and its outcomes. However, the high resource requirements and costs associated with training LTGMs have made them practically impossible for individual artists to train\footnote{For example, a recent work by Chen et al.~\cite{chen_2024_PixArtalphaFast} propose a model that can be trained with a fraction of the cost it takes to train a large model such as Stable Diffusion V1.5. However this fraction amounts to \$26,000.}, and replaced process ownership with a contested product ownership~\cite{epstein_2023_ArtScienceGenerative}. In this paper, we map out the landscape of options available to artists for creating personal artworks, on LTGMs or personally-trainable models, and argue for revisiting model crafting as a significant venue for artistic personalization.
	
	\section{The Landscape of Personalization}
	In a 2018 interview~\cite{zachariou_2018_MachineLearningArt}, computational artist Memo Akten outlined a spectrum of approaches for artistic creation with deep generative models along two dimensions: data (creating one's own, curating, or reusing existing sets) and models (designing original algorithms, modifying existing ones, or using pre-trained models). Having tried every combination, Akten argued that as artists move from using a custom to existing models/data in their work, it becomes "harder to give it a unique spin and make it your own". Inspired by this analysis, we explore the current landscape of options that artists have for creating personal artworks, with findings detailed in \autoref{table:landscape} and discussed later. This landscape of options expands on the "spectrum of working" by Ploin et al.~\cite{ploin_2022_AIArtshow}, and emphasizes the degree of artistic control and the systems available to artists on different parts of the landscape.
	
\begin{table}[]
	\centering
	\caption{Finer creative control over generative models increase as we move (down) between navigating, adapting or crafting models, it also increases as we move (left) from training on existing, curated or manually created datasets or as we move (left) from large to small datasets.}
	\label{table:landscape}
	\begin{tabular}{|c|c|ccc|cc|cc|c|}
		\hline
		\multicolumn{1}{|l|}{\multirow{2}{*}{}} &
		\multirow{2}{*}{\begin{tabular}[c]{@{}c@{}}\\\\Personalization\\ Method\end{tabular}} &
		\multicolumn{3}{c|}{\begin{tabular}[c]{@{}c@{}}Data \\ \hline Small <---> Large \end{tabular}} & 	\multicolumn{2}{c|}{\begin{tabular}[c]{@{}c@{}}Parts\\Updated\end{tabular}} &
		\multicolumn{2}{c|}{\begin{tabular}[c]{@{}c@{}}Model\\Training\end{tabular}} &
		\multirow{2}{*}{\begin{tabular}[c]{@{}c@{}}\\\\Examples of\\ Systems Used\end{tabular}} 
		\\ \cline{3-9}  
		
		\multicolumn{1}{|l|}{} &
		&
		\multicolumn{1}{c|}{\rotatebox[origin=c]{90}{ Created }} &
		\multicolumn{1}{c|}{\rotatebox[origin=c]{90}{ Curated }} &
	
		\rotatebox[origin=c]{90}{ Existing }
		 &
		\multicolumn{1}{c|}{\rotatebox[origin=c]{90}{Weights}} &
		\rotatebox[origin=c]{90}{Network} &
		\multicolumn{1}{c|}{\rotatebox[origin=c]{90}{ From Scratch }} &
		\rotatebox[origin=c]{90}{ Pre-trained } & \\ \hline 
		\rotatebox[origin=c]{90}{ Navigate } &
		\begin{tabular}[c]{@{}c@{}}Prompting, sampling,\\ interpolating, ..etc\end{tabular} &
			\multicolumn{3}{c|}{\multirow{7}{*}{
					\begin{tabular}[t]{l}moving here\\ down ($\downarrow$) \\or left ($\leftarrow$)\\ increases\\creative\\control\\  and needed \\effort\end{tabular}}} &
		\multicolumn{1}{c|}{\xmark} &
		\xmark &
		\multicolumn{1}{c|}{\xmark} &
		\cmark &
		\begin{tabular}[c]{@{}l@{}}Img2img~\cite{2024_Automatic1111WebUI} or Latent Projection~\cite{2024_AutolumeNeuralnetworkbased}\end{tabular} \\ \cline{1-2} \cline{6-10} 
		\multirow{3}{*}{\rotatebox[origin=c]{90}{ Adapt }} &
		Few-shot model adaptation &
		\multicolumn{3}{c|}{} &
		\multicolumn{1}{c|}{\cmark} &
		\xmark &
		\multicolumn{1}{c|}{\xmark} &
		\cmark &
		\begin{tabular}[c]{@{}l@{}}Textual Inversion~\cite{2024_Automatic1111WebUI, 2024_ComfyAI, 2024_Photobooth}\end{tabular} \\ \cline{2-2} \cline{6-10} 
		&
		Model fine-tuning &
		\multicolumn{3}{c|}{} &
		\multicolumn{1}{c|}{\cmark} &
		\xmark &
		\multicolumn{1}{c|}{\xmark} &
		\cmark &
		\begin{tabular}[c]{@{}l@{}}Tuning a pre-trained model~\cite{2024_AutolumeNeuralnetworkbased, 2024_RunwayML}\end{tabular} \\ \cline{2-2} \cline{6-10} 
		&
		Train an off-the-shelf model &
		\multicolumn{3}{c|}{} &
		\multicolumn{1}{c|}{\cmark} &
		\xmark &
		\multicolumn{1}{c|}{\cmark} &
		\xmark &
		\begin{tabular}[c]{@{}l@{}}Picking a GAN variation~\cite{2024_AutolumeNeuralnetworkbased, 2024_RunwayML}\end{tabular} \\ \cline{1-2} \cline{6-10} 
		\multirow{3}{*}{\rotatebox[origin=c]{90}{Craft}} &
		Active Divergence~\cite{broad_2021_Activedivergencegenerative} &
		\multicolumn{3}{c|}{} &
		\multicolumn{1}{c|}{\cmark} &
		\cmark &
		\multicolumn{1}{c|}{\cmark} &
		\cmark &
		Network Blending~\cite{2024_AutolumeNeuralnetworkbased} \\ \cline{2-2} \cline{6-10} 
		&
		\begin{tabular}[c]{@{}c@{}}Train a modified model\end{tabular} &
		\multicolumn{3}{c|}{} &
		\multicolumn{1}{c|}{\cmark} &
		\cmark &
		\multicolumn{1}{c|}{\cmark} &
		\xmark &
		\begin{tabular}[c]{@{}l@{}}Customization frameworks~\cite{sankaran_2018_IBMGANToolkit, pal_2019_TorchGANflexibleframework}\end{tabular} \\ \cline{2-2} \cline{6-10} 
		&
		\begin{tabular}[c]{@{}c@{}}Build own model\end{tabular} &
		\multicolumn{3}{c|}{} &
		\multicolumn{1}{c|}{\cmark} &
		\cmark &
		\multicolumn{1}{c|}{\cmark} &
		\xmark &
		\begin{tabular}[c]{@{}l@{}}Deep Learning libraries~\cite{paszke_2019_PyTorchimperativestyle, abadi_2015_TensorFlowLargescalemachine}\end{tabular} \\ \hline
	\end{tabular}
\end{table}

	\subsection{Navigating Generative Spaces}
	By interacting with a generative model, artists navigate the generative space of the model, i.e. its space of possibilities, to find aesthetics that interest them. Navigation can take place through prompting (e.g., ~\cite{openai_2024_DALLE, 2024_MidJourney}), latent projection and semantic sliders (e.g., on Autolume~\cite{2024_AutolumeNeuralnetworkbased}), canvas-like~\cite{rost_2023_Stablewalkinteractive} and paining-like~\cite{chung_2023_PromptPaintSteeringtexttoimage} interactions, or by embedding generative models into bespoke workflows~\cite{2024_ComfyAI, derivative_2024_TouchDesignernodebasedvisual}, among others~\cite{shi_2023_HCICentricsurveytaxonomy}. Steering interfaces that provide semantically meaningful controls improve artists' control, self-efficacy, and creative ownership~\cite{louie_2020_NoviceAIMusicCoCreationa}. We contend that navigation becomes more effective when it takes place in the generative space of expressive~\cite{louie_2022_ExpressivecommunicationEvaluatingb} and personalized models~\cite{abuzuraiq_2024_PersonalizingGenerativeAI}. 
	
	\subsection{Adapting Generative Spaces}
	When navigating the model's generative space proves cumbersome or if it consistently fails to produce personal results, artists can adapt generative models, which helps in shaping and focusing the scope of navigation. Few-shot model adaptation relies on a handful of example images and can introduce novel concepts into the generative space of the model with minimal changes. Full fine-tuning (and training from scratch) requires more example images but it can produce visual results for domains that are difficult to model with a few images such as Arabic Calligraphy~\cite{sobhansarbandi_2021_NavigatinglatentExploring}. Numerous techniques for adapting and personalizing generative models exist~\cite{robb_2020_Fewshotadaptationgenerative, hu_2021_LoraLowrankadaptation, gal_2022_imageworthone, ruiz_2023_DreamboothFinetuning}. Furthermore, visual arts creation tools (e.g.,~\cite{2024_Automatic1111WebUI, 2024_ComfyAI, 2024_InvokeAIcommunityversion, 2024_InvokeAIindustryversion, 2024_RunwayML, 2024_PlayformNoCodeAI, 2024_AutolumeNeuralnetworkbased}) often integrate adaption into their interfaces, whether they operate on personally-trainable or large generative models, such that the adapted models become available for downstream navigation~\cite{abuzuraiq_2024_PersonalizingGenerativeAI}
	
	\subsection{Crafting Generative Spaces}
\begin{figure}[]
		\centering
		\includegraphics[width=0.7\linewidth]{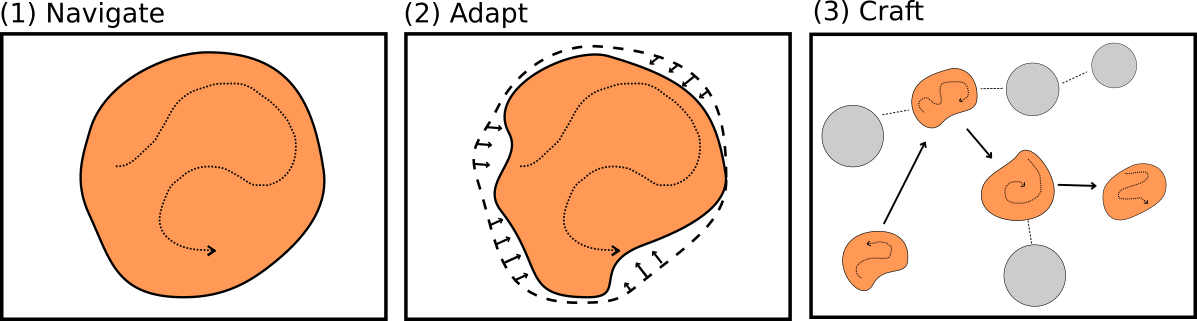}
		\caption{Three modes for creating personal works using generative models: (1) model navigation, (2) model adaption, and (3) model crafting. In the first mode, artists navigate within a fixed generative space to find the aesthetics they like, e.g. by prompting. In the second, navigation takes places within an adapted generative space. In the third, artists (co-)explore the space of generative models (straight arrows) as well as navigating within the generative space of each model (curved arrows).}
		
		\label{fig:three_modes}
	\end{figure}

	During model navigation a generative model is not modified, instead artists only sample from its generative space (\autoref{fig:three_modes} - 1). During model adaption, the weights of the model are changed which results in a different generative space to navigate in (\autoref{fig:three_modes} - 2). On the other hand, model crafting involves manipulating the model's architecture including designing new architectures, or editing and mixing existing models (c.f. in~\hyperref[sec:active_divergence]{Active Divergence}), which can produce qualitatively different and more personal generative spaces than is possible with model adaption alone, by virtue of affording finer control. Conceptually, model crafting can be defined as a creative exploration in the design space of generative models (\autoref{fig:three_modes} - 3). Given the high computational cost of training large models, model crafting for artists is limited to small models in practice, noting that "small" depends on the artist's access to computational resources.

	\subsubsection{Active Divergence}
	\label{sec:active_divergence}
	Artists often seek novelty and surprise from generative models. However, generative deep learning models are, by definition, trained to model a target distribution. Active divergence methods~\cite{broad_2021_Activedivergencegenerative}, as the name indicates, are a collection of methods that diverge generative models from a target distribution to produce novel and out-of-distribution samples. Active divergence methods, such as network rewriting~\cite{bau_2020_Rewritingdeepgenerative}, bending~\cite{broad_2021_NetworkbendingExpressive}, and blending~\cite{pinkney_2020_Resolutiondependentgan}, involve re-purposing and manipulating pre-trained models for art creation, with or without data.
	
	\subsubsection{Model Crafting vs. Model Development} Developing generative models, and machine learning models in general, is a mature field with support tools that span the entire development lifecycle including data preparation and pre-processing, hyperparameters optimization, model version control, and training process tracking as well as tools for testing and deploying models~\cite{budras_2022_indepthcomparisonexperiment, vidaldominguez_2023_ToolsDeepLearning}. Artists can lean into this ecosystem in developing generative models for creative applications, but the emphasis on efficiency, generality, and scalability, around which those tools are built is not always relevant for art creation. In fact, when it comes to art creation, the bias and overfitting associated with small datasets or simple models can be desired~\cite{vigliensoni_2022_smalldatamindsetgenerative}. Art creation involves experimentation, visual exploration, and reflection and is characterized by the lack of optimal solutions, which is why we prefer the term Model Crafting over Model Development. If we concede that crafting for creative endeavors is different from model development for common purposes, we can deduce that it also requires different tool support. As the system examples in \autoref{table:landscape} and previous sections show, model navigation and adaption for the visual arts are well supported with no-code tools while crafting largely happens on coding editors and paper sketches~\cite{wongsuphasawat_2017_Visualizingdataflowgraphs}. A notable exception is Autolume~\cite{2024_AutolumeNeuralnetworkbased, kraasch_2022_AutolumeLiveTurningGANs}, a no-code visual synthesis system that combines model navigation (via semantic sliders and latent projection), adaption ( fine-tuning), and crafting (via network blending).
	
	\section{Designing for Model Crafting}
		\textbf{3.1. Challenges} Model crafters and researchers who design tools for them face multiple challenges including: (1) the prolonged training times for training generative models, and (2) the interdisciplinary and technical expertise required for model crafting. However, we join Abuzuraiq and Pasquier~\cite{abuzuraiq_2024_PersonalizingGenerativeAI} in arguing that the proliferation of cloud computing and the advances on data-efficient, limited-data and few-shot generative models~\cite{abdollahzadeh_2023_surveygenerativemodeling,li_2022_comprehensivesurveydataefficient,yang_2023_Imagesynthesislimited,moon_2022_Finetuningdiffusionmodels} can lead high-quality models that are faster to train (technical-perspective). Furthermore, that the adoption of "small data"~\cite{vigliensoni_2022_smalldatamindsetgenerative} and "slow technology"~\cite{hallnas_2001_Slowtechnologydesigning} mindsets can provide effective lenses for the design of model crafting tools (design-perspective).
		
		\textbf{3.2. Opportunities} Large generative AI systems impact the artist’s ability to perceive generated works as their own (i.e. authorship), and the systems' black-boxed nature impacts the artist's sense of control over the results (i.e., agency). In this context, model crafting offers the following distinct advantages:
	
	\textbf{A. Owning both the Creative Process and its Products:} Creative AI, epitomized by recent generative AI systems, emphasizes the products over processes~\cite{colton_2019_computationalcreativitycreative}. When people are presented with a generative AI system that creates impressive results without offering them means to understand or contribute to its process, their agency is decreased and they lose an opportunity to reflect on and own their creative process, in addition to peering into human and computational creativity at large. The advent of large text-to-image generative models raised questions over the attribution (authorship) of what is produced and challenged artist's sense of ownership (agency) when using those models, possibly due to the incurred loss of control~\cite{steinbruck_2023_Creativeownershipcontrol}. To reclaim agency over the created products, artists personalize and adapt generative models. However, product ownership (and authorship) can also come as a by-product of process ownership, which can be achieved through model crafting along with carefully curated or created datasets. 
	
	\textbf{B. Sense of Craftsmanship:} Model crafting is challenging and requires weaving expertise that spans coding, deep learning, and specific domain expertise. But it is also a rewarding activity that can be a source of pride and craftsmanship, which is a motivation they share with makers in the Do-It-Yourself (DIY) community~\cite{milne_2014_Whatmakesmaker}. Crafters of generative systems also often report surprising~\cite{elgammal_2023_TexttoImageGeneratorsHave} and serendipitous~\cite{lehman_2020_surprisingcreativitydigital} encounters as they experiment with generative systems. 

	\textbf{3.3. Creativity Support for Model Crafting:}  Given the advantages outlined above, we encourage researchers to study and design tools for the model crafters. The work in this direction is scarce, so we weave together some starting points. First, by analogy to sketching in design, Lam et al~\cite{lam_2023_ModelSketchingCentering}, propose a system for Model Sketching by which users explore high-level concepts (e.g. profanity or racism) based on which ML classifiers are subsequently created. In the same vein of simplifying ML model design for non-experts, multiple visual programming frameworks for deep learning are introduced~\cite{tamilselvam_2019_visualprogrammingparadigma, calo_2023_visualprogrammingtool}, as well as direct manipulation tools for visualizing and experimenting with deep learning models~\cite{smilkov_2016_DirectmanipulationVisualizationDeep}. To reduce the barrier for non-expert users, Croisdale et al.~\cite{croisdale_2023_DeckFlowCardGamea} propose a data-flow system of cards for exploring multi-modal generative workflows. Similarly, Compton et al.~\cite{compton_2017_generativeframeworkgenerativity} present physical cards for creating generative pipelines. Finally, crafting involves finer control including mixing and matching. Ulberg et al.~\cite{ulberg_2020_Handcraftingneuralnetworks} suggest visualizing and crafting model weights for better control over neural networks for the arts. Shimizu et al.~\cite{shimizu_2023_Interactivemachinelearning} enable creating bespoke text-to-media mappings between generative models with a few examples.
	
	\section{Conclusion \& Future Work}
	Crafting, adapting and using generative systems has a long and diverse lineage in many academic fields and communities of practice, even if not under the term Generative AI or if not based on machine learning~\footnote{Generative systems can be rule-based, probabilistic or constraint satisfaction systems, or they can use genetic search algorithms, and machine learning models (specifically generative deep learning)}. Other names included procedural content generation in games, generative design in architectural or industrial design, generative music, and computational or generative art to name a few. Through model crafting, artists can "hack" or subvert generative AI systems, allowing them to create critical statements about AI systems and how they relate to society~\cite{ploin_2022_AIArtshow}. Such inquiries into the inner working of AI systems have resulted in pioneering works such as DeepDream~\cite{mordvintsev_2015_InceptionismGoingdeeper}, a predecessor on visualizing neural networks' activations. Finally, the work on creative model crafting for personally-trainable models can inform the design and development of large generative models as well. In the future, we will continue to design for and study model crafting as a creative endeavor in its own right.
	% With roots in extensive bodies of work on generative and procedural systems, model crafting can also extend its reach to other fields. 

	\bibliographystyle{ACM-Reference-Format}
	\bibliography{main.bib}

\end{document}